\documentclass[twocolumn]{aastex62}

\usepackage[utf8]{inputenc}
\usepackage[english]{babel}
\usepackage[T1]{fontenc}
\usepackage{ mathrsfs }
\usepackage{amsmath}
\usepackage{natbib}

\received{January 14, 2020}
\revised{February 27, 2020}
\accepted{February 28, 2020 for publication in The Astrophysical Journal}

\begin{document}

\title{Coronal Bright Points as possible sources of density variations in the Solar Corona}

\author{Léa Griton}
\email{lea.griton@irap.omp.eu}
\affiliation{IRAP, Universit\'e Toulouse III - Paul Sabatier,
CNRS, CNES, Toulouse, France}

\author{Rui F. Pinto}
\affiliation{IRAP, Universit\'e Toulouse III - Paul Sabatier,
CNRS, CNES, Toulouse, France}

\author{Nicolas Poirier}
\affiliation{IRAP, Universit\'e Toulouse III - Paul Sabatier,
CNRS, CNES, Toulouse, France}

\author{Athanasios Kouloumvakos}
\affiliation{IRAP, Universit\'e Toulouse III - Paul Sabatier,
CNRS, CNES, Toulouse, France}

\author{Michäel Lavarra}
\affiliation{IRAP, Universit\'e Toulouse III - Paul Sabatier,
CNRS, CNES, Toulouse, France}

\author{Alexis P. Rouillard}
\affiliation{IRAP, Universit\'e Toulouse III - Paul Sabatier,
CNRS, CNES, Toulouse, France}

\begin{abstract}
Recent analysis of high-cadence white-light images taken by the Solar-Terrestrial RElations Observatory (STEREO) near solar maximum has revealed that outflowing density structures are released in an ubiquitous manner in the solar wind. The present study investigates whether these density fluctuations could originate from the transient heating of the low corona observed during Coronal Bright Points (CBPs). We assume that part of the intense heating measured during CBPs occurs at the coronal base of open magnetic fields that channel the forming solar wind. We employ the solar wind model MULTI-VP to quantify the plasma compression induced by transient heating and investigate how the induced perturbation propagates to the upper corona. We show that for heating rates with statistics comparable to those observed during CBPs the compressive wave initially increases the local plasma density by a factor of up to 50$\%$ at 5 R$_\odot$. The wave expands rapidly beyond 30 solar radii and the local enhancement in density decreases beyond. Based on the occurrence rates of CBPs measured in previous studies, we impose transient heating events at the base of thousands of open magnetic field lines to study the response of the entire 3-D corona. The simulated density cubes are then converted into synthetic white-light imagery. We show that the resulting brightness variations occupy all position angles in the images on timescales of hours. We conclude that a significant part of the ubiquitous brightness variability of the solar corona could originate in the strong transient heating of flux tubes induced by CBPs.

\end{abstract}


\section{Introduction}
\label{sec:intro}

The past two decades of coronal observations have revealed the highly variable nature of the solar corona on all measurable spatial and temporal scales from the deep layers of the solar atmosphere to the upper corona. The relation of coronal variability to fundamental processes such as coronal heating, the formation of the solar wind or the acceleration of energetic particles is still unclear and remains a topic of active research.

The recent analysis of high-cadence white-light observations made by the COR-2 instrument \citep{Howard2008} onboard the Solar-Terrestrial RElations Observatory \citep{Kaiser2008} during a dedicated deep-field campaign \citep{Deforest2018} reveals that brightness variations appear at all latitudes from the corona, superposed on a thinly striated background corona.
The observation campaign occurred around April 14, 2014, during solar maximum but in the absence of any major Coronal Mass Ejection (CME). After a precise study of all possible errors due to the data treatment, they inferred that the variability observed in the images is due to in-homogeneous density structures flowing out of the corona, with fluctuations of the order of possibly ten times the average density. Figure 12 of \citet{Deforest2018} shows clearly that brightness fluctuations are present at all latitudes and times, covering different length scales and intensities. Since the brightness of the corona strongly depends on electron density, density fluctuations may exist in all types of solar winds between 5 and 15 solar radii ($R_\odot$). 

The largest structures observed by \citet{Deforest2018} were related in the study to streamer blobs released in the slow solar wind from the tip of streamers \citep{Sheeley1997}. Many of these structures have been related in past studies with the release of magnetic flux ropes \citep{Sheeley2009} measured frequently near sector boundaries in the solar wind \citep{Rouillard2009,Sanchez-Diaz2017a, Sanchez-Diaz2017b} (for previous studies, please see references therein). These past studies have found that blobs form by magnetic reconnection near the tip of helmet streamers at 3 to 5 solar radii (R$_\odot$).\\

The ubiquitous small-scale variability highlighted in the \citet{Deforest2018} is, however, unlikely to be resulting only from transient events near the tip of streamers. By simple visual inspection, \citet{Deforest2018} point out that at least one of the largest density structures could originate in the low solar atmosphere as a coronal bright point (CBP). CBPs are localized increases in extreme-ultraviolet (EUV) and X-ray emissions that are detected in the low corona \citep{Madjarska2019}. They often appear as hot loops (log$T$[K]$\simeq 6.2$) overlying cooler ones (log$T$[K]$\simeq 6$) with cooler legs (log$T$[K]$\simeq 4.9$) \citep{Kwon2012}. The combined analysis of intense CBPs observed in EUV and white-light images of the upper corona have revealed that outflowing density structures known as 'coronal jets' often originate as CBPs low in the corona \citep{Wang1998}. Recent numerical simulations also corroborate this view \citep[\emph{e.g,}][]{singh_modelling_2019}. These coronal jets become part of the outflowing solar wind outstreaming from coronal holes \citep{Wang1998}. Hence an association exists between the occurrence of density structures in the outflowing solar wind and the occurrence of CBPs low in the corona. A recent analysis of equatorial coronal holes has revealed the presence of CBPs at the footpoints of open magnetic flux tubes that appears to influence strongly the occurrence of coronal plumes. Intense heating \citep{Raouafi2014} and dynamical perturbations \citep{pinto_coronal_2010} can lead to a sustainable enhancement of plasma density that could contribute to the development of coronal plumes.
Previous observation campaigns made with LASCO \citep{cho_two-dimensional_2018} and UVCS \citep{bemporad_exploring_2017} furthermore highlight the existence of a highly structured background solar wind all the way from the low to the high corona (and at different epochs), visible both in plasma density and outflow speed.
\\

The statistical analysis of CBPs by \citet{Alipour2015} reveals that, in April 2014, when the deep-field campaign of \cite{Deforest2018} took place, the average daily number of CBPs observed on the entire visible solar disk was around 550. If one extrapolates this number to the full photosphere, covering 41253 $deg^2$ in total, we can infer that in April 2014 there were about 0.02 CBP per $deg^2$ per day. Since white-light coronagraphs integrate brightness variations over an extended region along a particular line of sight, outflowing structures observed along a particular position angle (or 'projected latitude') can originate from a broad range of longitudes situated well in front and behind the plane of sky. A coronagraph is sensitive to brightness variations situated anywhere inside an angular distance of approximately 40 degrees on either side of the plane of the sky. Therefore we would expect 1.6 CBPs to occur over this region in the lower corona for each one degree-band of position angle. If each of these CBPs were equally spread out in time and somehow led to a density variation that was sufficiently strong to be detected in white-light images, this would produce an apparent periodic brightness variation of 15 hours. \\ 

The goal of our study is to investigate whether transient heating processes with properties similar to CBPs occurring low in the corona can lead to the type of density structures observed in the corona by STEREO in the solar wind. In order to do so, we ran a series of solar wind simulations that take into account the specific magnetic topology of the corona of the \citet{Deforest2018} deep-field campaign, to which we introduce perturbations that mimic localised heating events caused by CBPs.\\

We begin by detailing the frame and the steps of our method (Sect. \ref{sec_methods}), first describing the simulation code and the equations. 
We then describe in section \ref{sec_results} the results of 1-D simulations (sub-sections \ref{sec_results_1Dref}, \ref{sec_results_1Dsh}, \ref{sec_results_1Dph}) that we ran to explore in detail the physical processes at play before analysing the 3-D simulations (sub-sections \ref{sec_results_3Dsh} and \ref{sec_results_3Drph}). The last section is dedicated to a general discussion about the simulations results in the context of recent observational work and provides a summary of the new conclusions.\\

\section{Method}{\label{sec_methods}}

\subsection{Solar wind model}

Transient heating events in the low corona are likely to cause different types of perturbations to the wind flows in their formation region. Several effects on the background solar wind can be foreseen, including, for example, direct changes to the thermal stratification of the corona, to the mass flux, and triggering of waves and shocks. The relative significance of such effects should depend on the characteristics of the heat deposition events (e.g, relative position, amplitude, impulsiveness and duration), on the specific state of the unperturbed solar winds streams, and on the geometry of the magnetic flux tubes that they flow along.
We chose to use the MULTI-VP code \citep{Pinto2017}, that computes full solar wind solutions (surface to high corona) taking into account heating and cooling processes in the corona.

MULTI-VP computes the profiles of a large ensemble of contiguous uni-dimensional solar wind streams from the surface of the Sun up to about $30\ R_{\odot}$. The individual streams are guided along individual magnetic flux tubes, whose geometry can either be idealized or, more commonly, obtained by means of magnetogram extrapolation and thus fully sample the three-dimensional corona.
The code solves a system of equations describing the heating and acceleration of each individual wind stream, guided along a given magnetic flux tube
\begin{eqnarray}
  \partial_t \rho  &+& \nabla\cdot\left(\rho\mathsf{u}\right)=0\ , \label{eq:rho} \\
  \partial_t \mathsf{u}&+&\left(\mathsf{u}\cdot\nabla_s \right)\mathsf{u} = -\frac{\nabla_s P}{\rho}  \nonumber \\
  &-& \frac{GM}{r^2}\cos\left(\alpha\right) + \nu\nabla_s^2u\ \label{eq:u}, \\
  \partial_t T &+& \mathsf{u}\cdot\nabla_s T + \left(\gamma-1\right)T\nabla\cdot\mathsf{u} =  \nonumber \\
  &-& \frac{\gamma-1}{\rho}\left[\nabla\cdot F_h + \nabla\cdot F_c +
    \rho^2\Lambda\left(T\right) \right] \label{eq:t}\ ,
\end{eqnarray}
where $\rho$ is the mass density, $\mathsf{u}$ is the wind speed, and $T$ is the plasma temperature.
The wind profiles are computed on a grid of points aligned with the magnetic field (with curvilinear coordinate $s$), $\alpha$ is the angle between the magnetic field and the vertical direction,
and $r$ represents the radial coordinate (distance to the center of the Sun).
The divergence operator is defined as
\begin{equation}
  \nabla\cdot\left(*\right) = \frac{1}{A\left(s\right)}\frac{\partial}{\partial s}\left(A\left(s\right) * \right) =
  B \frac{\partial}{\partial s}\left(\frac{*}{B}\right) \ ,
\end{equation}
where $B\left(s\right) \propto 1/A\left(s\right)$, $A\left(s\right)$ being the flux tube's cross sectional area and $B\left(s\right)$ the magnetic field amplitude.
The ratio of specific heats is $\gamma = 5/3$.
The terms $F_{\rm h}$, $F_{\rm c}$ denote the mechanical heating flux and the
Spitzer-Härm conductive heat flux, which are both field-aligned.
The radiative loss rate is $\Lambda\left(T\right)$.

For simplicity, we associate here the mechanical heating flux $F_{\rm h}$ (or, equivalently, the heating rate $Q_h = -\nabla\cdot F_h$) to a function that represents the effects of the coronal heating processes (rather than including the actual small-scale dissipation processes)
\begin{equation}
  \label{eq_flux_global}
  F_{\rm h} = F_{\rm B0} \left(\frac{A_0}{A}\right)h(s). 
\end{equation}
The coefficient $F_{\rm B0}$ is proportional to the basal magnetic field amplitude $\left|B_0\right|$ (with, for basal field amplitudes consistent with typical Wilcox Solar Observatory magnetograms, $F_{\rm B0}=12 \times 10^5 \left|B_0\right|\ \mathrm{erg\ cm^{-2}\ s^{-1}}$).
The coronal heating function $h\left(s\right)$ is split into two terms, the first one corresponding to a permanent background heating
\begin{equation}
  \label{eq_fluxp}
  h_b\left(s\right)=\exp\left[-\frac{s-R_\odot}{H_{\rm f}}\right].
\end{equation}
The damping scale-height $H_{\rm f}$ is anti-correlated with the superradial expansion ratio $f_{SS}$ in the low corona, as in \citet{Pinto2017}.
The second heating term corresponds to an additional time-dependent heat source confined to the low corona.
We defined this function such that the corresponding heating rate $Q_{ht}$ assumes a Gaussian profile in respect to height.
The total heating rate then becomes
\begin{equation}
  \label{eq_heatingrate_exp}
  \begin{split} 
    Q_{\rm h}= & - F_{\rm B0} \left(\frac{A_0}{A}\right) \times \\ 
        & \left(\frac{h_b(s)}{H_{\rm f}} + \frac{1}{r_{\rm p}}\mathscr{H} a_{\rm p}\exp\left[-\frac{(s-R_{\rm p})^2}{r^2_{\rm p}}\right] \right),
    \end{split}
\end{equation} 
where $R_{\rm p}$ is the radial position of the Gaussian function and $r_{\rm p}$ defines the width of the Gaussian function. The factor $a_{\rm p}$ lets us change the amplitude of the perturbation relatively to the main permanent heating term. The factor $\mathscr{H}\left(t\right)$ modulates the temporal dependence of the heating perturbation (letting us switch it on and off).
Our approach differs from (and complements) those of previous simulations of intermittent heating on solar wind flows by \citet{Lie-Svendsen2002,grappin1999} and \citet{pinto_time-dependent_2009} in many ways. 
The former used a higher-order multi-fluid model to investigate how time-dependent heating of protons could explain Ulysses' observations of slow and fast solar wind properties, especially velocity and temperature. The second studied periodic temperature variations due to impulsive event occurring below their actual numerical domain (a polytropic corona), and the latter studied extended (and more diffuse) heating sources that stretched from the chromosphere to the low corona.
Here, we focus on strong and compact heating sources well localised between the transition region (TR) and the first layers of the corona. We furthermore consider the full three-dimensional solar wind evolving on a magnetised medium whose geometry is constrained by magnetogram data.
We derive the three-dimensional structure of the coronal magnetic field from Potential Field Source-Surface extrapolations (PFSS) of a Wilcox Solar Observatory (WSO) synoptic magnetogram for Carrington rotation 2149, except for the 1-D simulations in Sects. \ref{sec_results_1Dref} to \ref{sec_results_1Dph}. The latter were based on idealized vertical flux tubes to better identify the effects of the heating perturbations on the wind flow.

\subsection{Simulation setup}

We ran a series of numerical simulations in order to identify clearly the effects of heating frequency, duration, position and width of the heating events on a simple and idealized geometry, before moving to a realistic three-dimensional case.
For all simulations, we prepared beforehand a stable solar wind solution that represented the background unperturbed flow (the reference case), that served as initial condition (see, e.g, Fig. \ref{fig_1}). The background wind simulations were setup in the same manner as the simulations in \citet{Pinto2017}. The main difference in simulation setup is that runs were performed at twice the radial resolution for the current study. The background heating parameters were kept unchanged, except for the magnetic field set-up of the test cases with a radially-aligned flux tube.
We then introduced perturbations to the background solar wind by adding localised and compact heat depositions (via the second term in the heating function in eq. \ref{eq_heatingrate_exp}) with different profiles and temporal dependence.

On a first set of simulations, we started from an idealized vertical and radially expanding magnetic flux tube, and we tested for different positions, widths and amplitudes of the heating perturbations (parameters $R_{\rm p}$, $r_{\rm p}$ and $a_{\rm p}$ in eq. \ref{eq_heatingrate_exp}). We started by making the heating event permanent. In other words, we defined the temporal factor $\mathscr{H}\left(t\right)$ as a Heaviside function

\begin{equation}
\label{eq_heaviside}
\mathscr{H}(t) =
        \left\{ \begin{array}{ll}
            1 & t>t_{\rm p} \\
            0 &\text{otherwise}
        \end{array} \right.
\end{equation}  

with $t_{\rm p}$ being the perturbation's starting time. These preliminary simulations helped us define boundaries for our parametric study, by eliminating set of parameters that would not lead to a new stationary state in a reasonable time (arbitrarily set to 15 hours of real time). Thus, we restricted our study to heating events occurring in the low corona, with no direct heat deposition at or below the TR. These could otherwise induce strong perturbations to the thermal equilibrium that determines its position and amplitude, and generate phenomena well beyond the scope of this study.
Subsequently, we re-ran the same simulations with a heating event with a finite duration, that is with

\begin{equation}
\mathscr{H}(t) =
        \left\{ \begin{array}{ll}
            1 & t_{\rm p} \leq t \leq t_{\rm p}+\delta t_1 \\
            0 &\text{otherwise}
            \label{eq_boxcar}
        \end{array} \right.
\end{equation}
with $\delta t_1$ representing the duration of the heating event.

Finally, we introduced recurrent heating events with a time period between consecutive events $\delta t_2$, such that
\begin{equation}
	\label{eq_periodic_boxcar}
	\begin{split}
	\mathrm{for}~t>  t_{\rm p} :\\
	\mathscr{H}(t) &= 1 ~\mathrm{ if }~ t-\delta t_2*floor(t/\delta t_2) \leq \delta t_1, \\
	~\mathrm{ else }~\mathscr{H}(t)  &= 0\ .
	\end{split}
\end{equation} 

We will discuss more thoroughly the dependence on the heating deposition height $R_{\rm p}$ than on the other parameters, as its variations provoke more varied effects (depending on the relation to the position of the transition region denoted $R_{\rm Tr}$, on the position of the temperature maximum $R_{\rm Tmax}$, and that of the sonic point). 
The heating parameters of the main set of one-dimensional simulations are listed in Table \ref{tab:parameters_1D}. \\

\begin{table}
    \centering
    \begin{tabular}{lcccc}
  \hline
   & $R_{\rm p} [R\odot]$ &  $\delta t_1$& $\delta t_2$ & 1D/3D\\
  \hline
   ref  & --      & --            &  --   & 1D \\
  1     & 1.2   & 12 min    &  --   & 1D\\
  2     & 1.2   & 1 min      & 1 h & 1D\\ 
  3     & 1.2   & 1 min      & --   & 1D\\
  4     & 1.42 & 1 min      & --   & 1D\\
  5     & 1.5   & 1 min      & --   & 1D\\
  6     & 1.2   & 5 sec       & --   & 1D\\
  7     & 1.2   & 2.25 min &  --  & 1D\\
  8     & 1.2   & 12 min    &  --  & 3D\\
  9     & 1.2   & random  & random & 3D\\
  \hline
    \end{tabular}
    \caption{Summary of the main heating parameters of the simulations discussed on this manuscript.}
    \label{tab:parameters_1D}
\end{table}

After analysing the 1-D simulations, we chose a preferred set of heating parameters, that we implemented on a sample of 1-D simulations (each representing an individual magnetic flux tube) filling the full three-dimensional sphere centered on the Sun, up to 30 solar radii.
The three-dimensional simulations are thus based on an ensemble of flux tubes that cover the whole open field as retrieved from PFSS extrapolations of WSO synoptic magnetograms for Carrington rotation 2149. This Carrington rotation covers the period studied  by \cite{Deforest2018}, and we hence try to provide a representation of the global magnetic field configuration around the Sun at the time of that study. 
We furthermore made both the heating duration and recurrence interval random, with mean values inspired on the results reported by \citet{Alipour2015} and \citet{Madjarska2019} for the same time period.

\section{Simulation results}{\label{sec_results}}

\subsection{Reference 1-D case}
\label{sec_results_1Dref}

\begin{figure}[!t]
\includegraphics[width=0.48\textwidth]{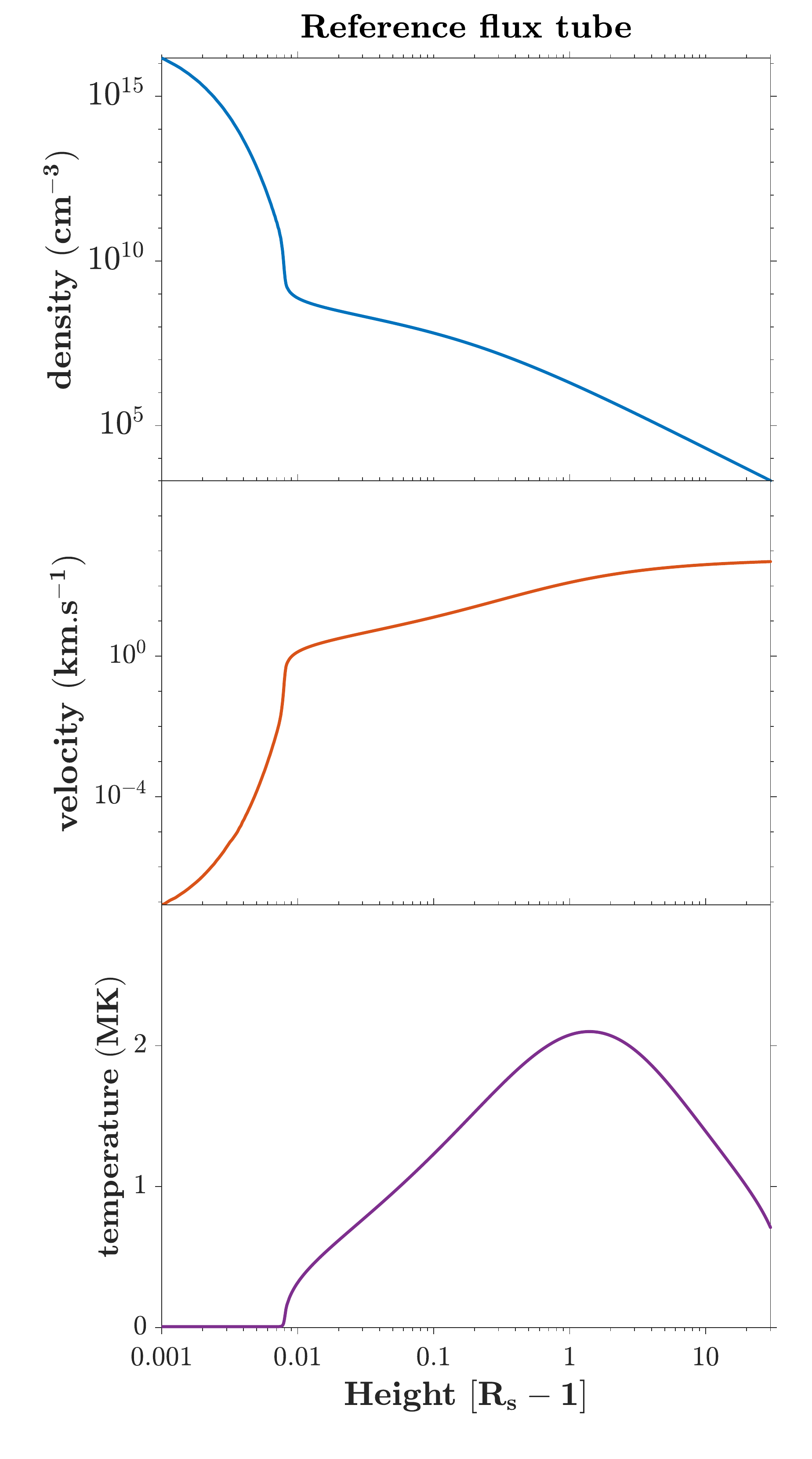}
\caption{{\small \textit{Simulation ``ref": Density, velocity, temperature and heating rate profiles of the initial equilibrium state, just before the change in Qh is introduced. }}} 
\label{fig_1}
\end{figure}

Figure \ref{fig_1} shows the radial profiles of the wind density, speed, temperature and heating rate per unit mass on our 1-D reference case (as a function of height above the solar surface). This case is set up with a vertical and radially expanding flux tube meant to be used as initial conditions for the subsequent 1-D simulations (Sects. \ref{sec_results_1Dsh}, \ref{sec_results_1Dph} and \ref{sec_results_ps}). The equilibrium state was achieved by letting the simulation run for more than 300 hours of physical time, until a perfectly stationary state was achieved. This time interval is much longer than the sound or slow mode transit time (the slowest modes allowed), that take approximately 10 hours to cross the entire simulation domain.

The reference case corresponds to a wind with a terminal speed of about $550\ \mathrm{km/s}$.
The transition region forms at about $R_{\rm Tr}=1.008\ R_{\odot}$, as a result of the thermal balance between the coronal heating rate, thermal conduction and radiative losses.
The plasma temperature peaks at $R_{\rm Tmax}=1.4\ R_{\odot}$.
The sonic point (where the wind speed becomes supersonic) is at $R_{\rm S}=2.48\ R_{\odot}$.

\subsection{Single heating event on a 1D wind stream}
{\label{sec_results_1Dsh}}

Reference case (Sect. \ref{sec_results_1Dref} and Fig. \ref{fig_1}) is perturbed with a single localised heating deposition (\emph{cf.} Eq. \ref{eq_heatingrate_exp}) with a finite duration. The additional heating term has the form of a Gaussian function with $0.01\ R_{\odot}$ width at half height and centered at $R_{\rm p}=1.2\ R_{\odot}$. The transient heating event has a duration of 12 minutes, in agreements with the mean duration of transient CBPs reported by \citet{Alipour2015}.
The width of the heating deposition region is of the same order as the maximum height generally observed for the loops at the origin of the CBPs \citep{Madjarska2019}. Its altitude was determined after a set of test simulations with varying heat deposition heights, that revealed that $R_{\rm p}=1.2 R\odot$ was optimal both in terms of producing clear pressure pulses in the low corona and of respecting the constraints of the numerical scheme.

\begin{figure}[!t]
\includegraphics[width=\linewidth]{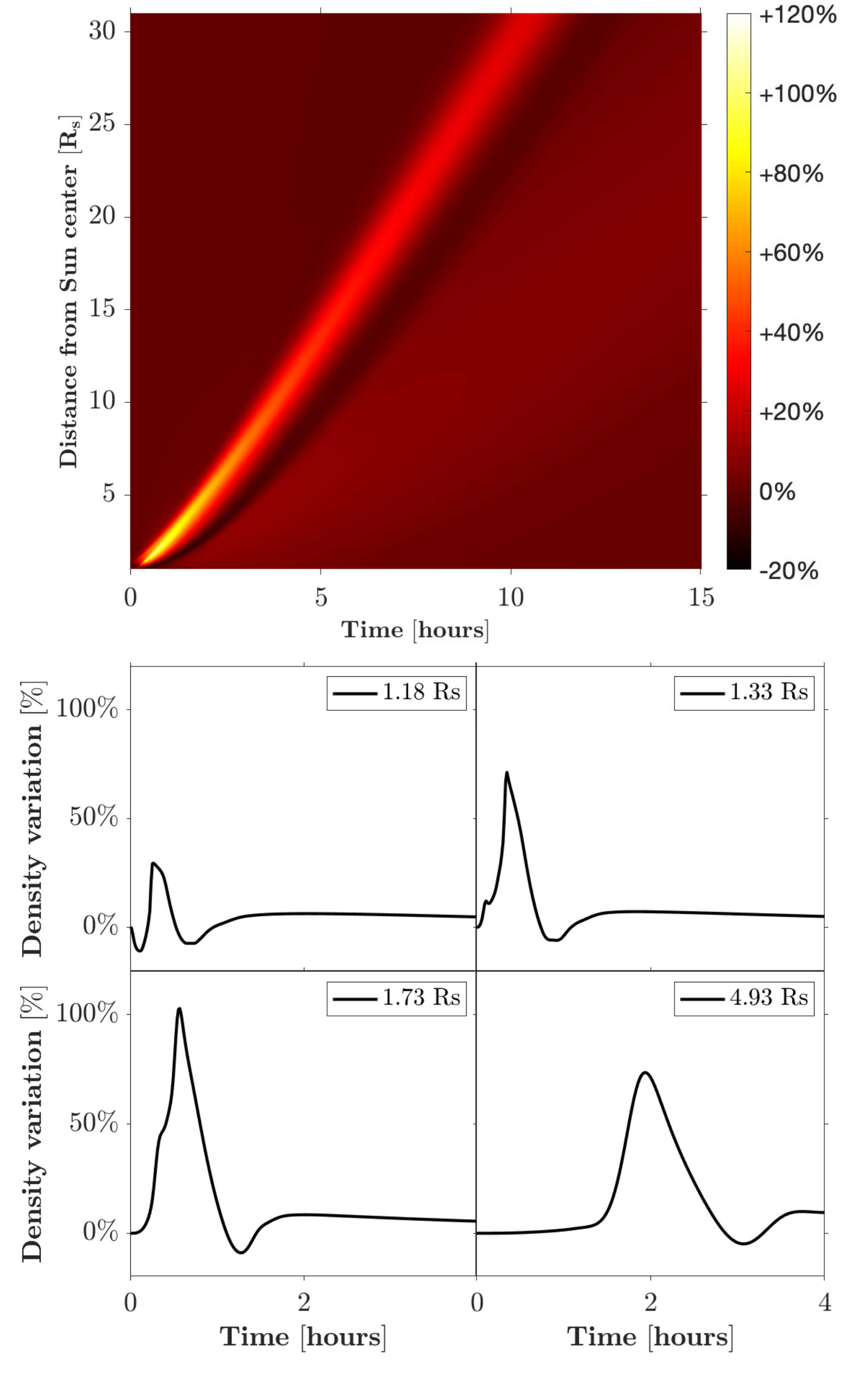}
\caption{{\small \textit{Simulation 1: Density variation along the entire simulation grid (from 1 to 31 $R\odot$), over 15 hours. One can notice the compression front travelling at the sound speed in the frame of the plasma (yellow) immediately followed by a less intense rarefaction front (black).}}} 
\label{fig_2}
\end{figure}

\begin{figure}[!t]
\includegraphics[width=\linewidth]{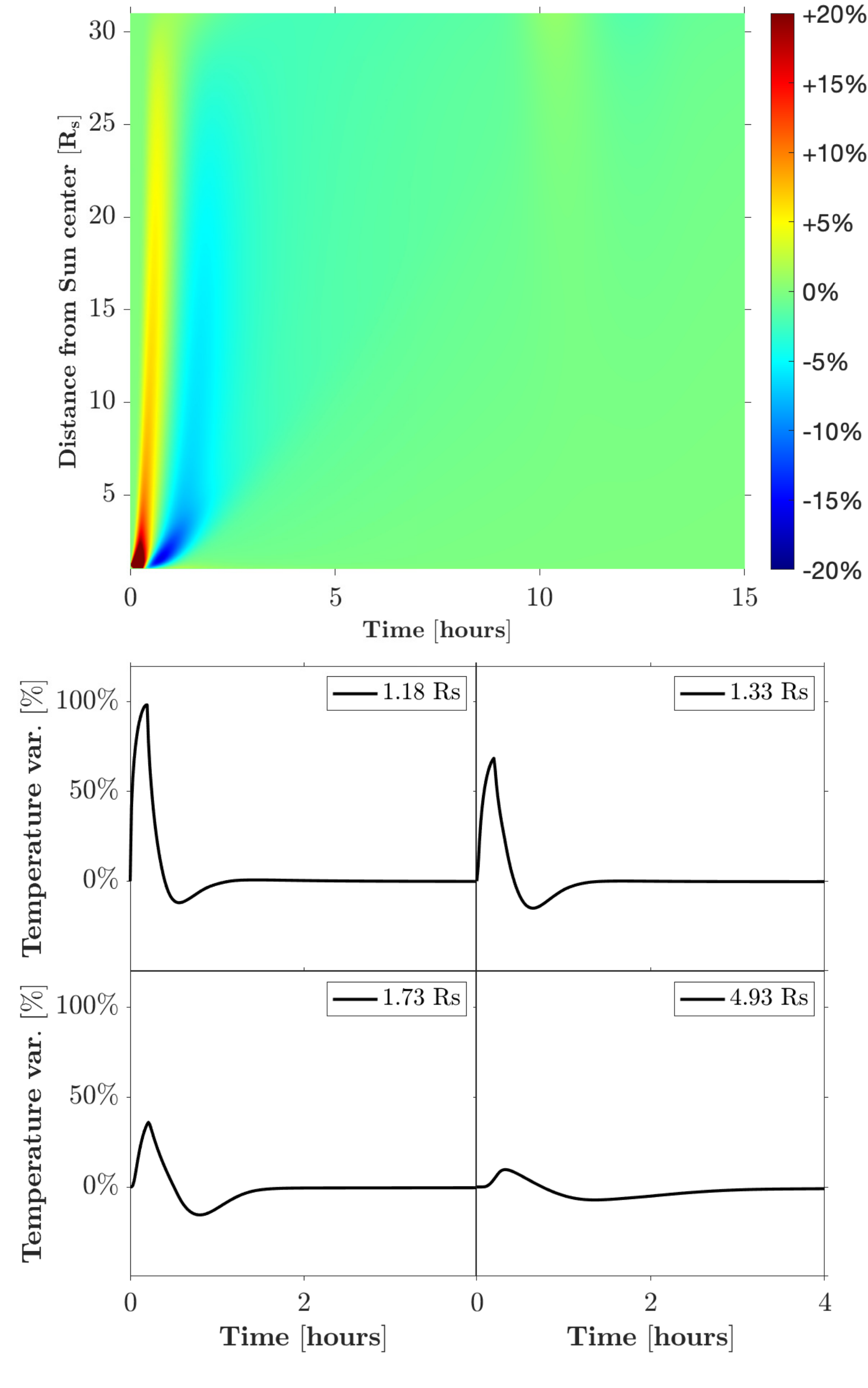}
\caption{{\small \textit{Simulation 1: Temperature variation along the entire simulation grid (from 1 to 31 $R\odot$), over 15 hours. One may note the change in the colour-bar values compared to the one on figure \ref{fig_2}, for the temperature is less intense than the density one.}}} 
\label{fig_3}
\end{figure}

Figures \ref{fig_2} and \ref{fig_3} show the evolution of the wind density and temperature following the 12 minute-heating event in the form of ``x-t" diagrams covering the whole flux tube length and the entire duration of the simulation (\emph{i.e.} 15 hours). The plots show the relative base-difference for each quantity $f$, computed as
\begin{equation}
  \label{eq_variation}
  df(t,r)= \frac{f(t,r)-f(0,r)}{f(0,r)},
\end{equation} 
with $f(0,r)$ being the value of the quantity $f$ at the position $r$ at the initial instant, before any additional heating is introduced.

The simulation shows that a compression front (yellow tones on figure \ref{fig_2}) forms just above the location of heating deposition ($R_p=1.2 R\odot$), and that it propagates upward along the flux tube to the upper boundary of the simulation grid. It is followed by a rarefaction front (with smaller relative amplitude) that  is generated at the end of the heating event.
The slope of the compression front in the ``x-t'' diagram corresponds to the sound speed in the frame moving with the plasma.

The sudden addition of a compact heating source immediately introduces a strong perturbation to the temperature profile. Thermal conduction is extremely efficient at coronal temperatures (as the conductive flux is $\propto T^{5/2}\nabla T$), and is the fastest energy transport mechanism available to the system.
Figure \ref{fig_3} shows the ``x-t'' diagram for the wind temperature and reveals the very quick propagation of a temperature enhancement, much quicker than the compression front on figure \ref{fig_2}. 

\begin{figure}[!t]
\includegraphics[width=\linewidth]{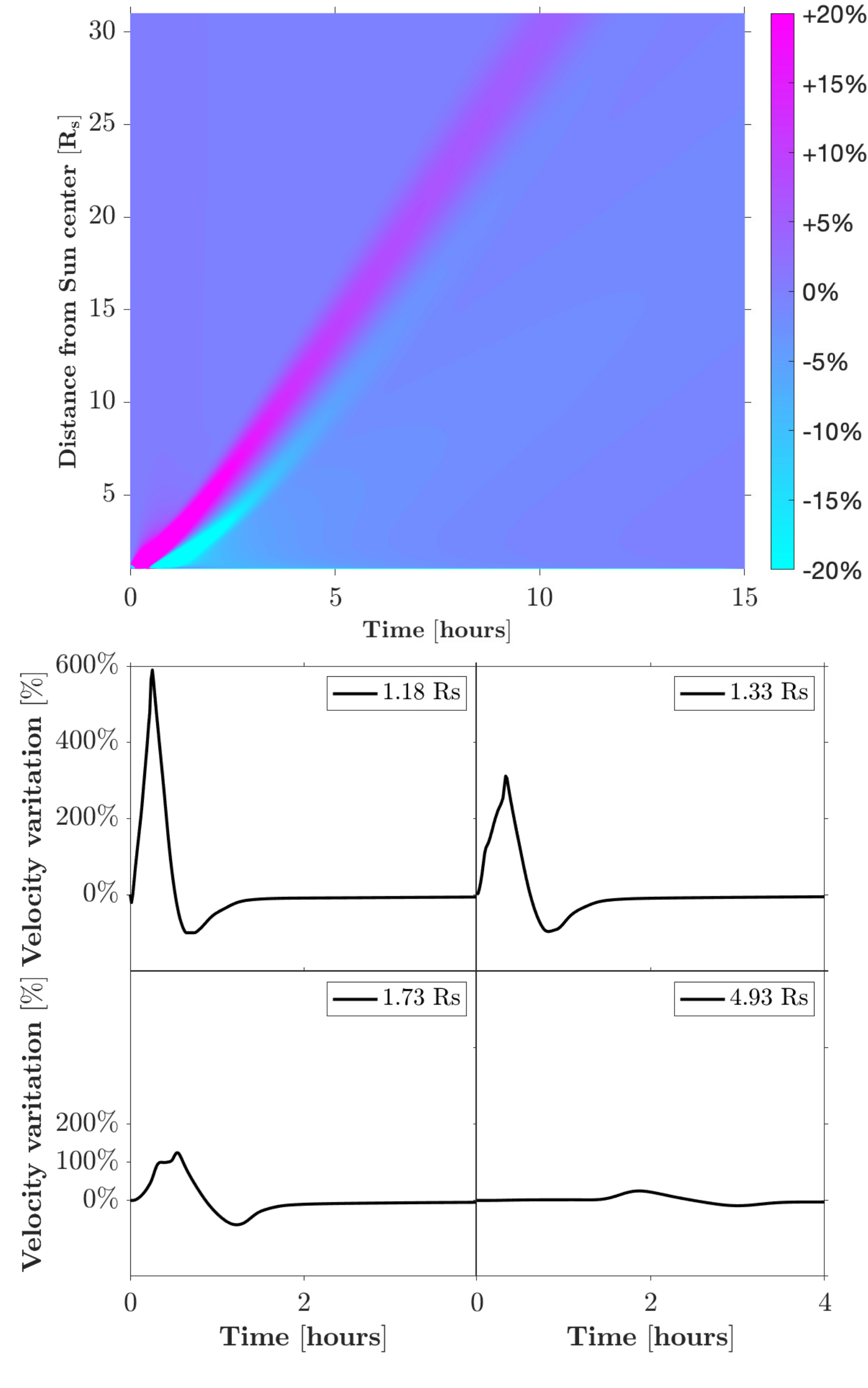}
\caption{{\small \textit{Simulation 1: Velocity variation along the entire simulation grid (from 1 to 31 $R\odot$), over 15 hours. One may note the change in the colour-bar values compared to the one on figure \ref{fig_2}, for the velocity variation is much more intense than the density one.}}} 
\label{fig_4}
\end{figure}

Figure \ref{fig_4} shows the ``x-t'' diagram for the wind speed perturbation, that reveals a propagation pattern that is nearly co-spatial with that of the density perturbation (with a positive $\delta v$ associated with a positive $\delta \rho$ for the compression front, and a negative $\delta v$ associated with a negative $\delta \rho$ for the rarefaction front, as is natural for forward propagating sound or slow-mode wavefronts).

The response to the heating perturbation presents a complex sub-structure.
The strong peak in temperature created by the sudden heat deposition results immediately in a set of positive and negative conductive heat fluxes that transport heat away from the heating location both outward and inward, respectively. The upward heat flux produces the strong positive $\delta T$ visible in Fig. \ref{fig_3}, while the downward negative flux drives a significant amount of energy into and across the transition region, which is then efficiently radiated away at the top of the chromosphere. The direct impact of the added conductive heat flux on the structure of the TR itself is small and short-lasting. However, the perturbations produce a longer-lasting increase of the nascent wind mass-flux (chromospheric mass evaporation) that is visible as a small positive $\delta \rho$ that follows the initial (stronger) compressive wavefronts and lingers on for a few hours.

The sudden heat deposition also produces a strong bi-directional dynamical response. The localised temperature increase acts as a pressure pulse that evacuates a significant amount of the deposited energy as compression fronts, propagating away from the perturbation both upward and downward. The upward propagating front leaves the domain after about 10 hours. The downward propagating front reaches the TR almost instantly. At this moment, a fraction of the wave front is reflected back into the corona (producing a weaker trailing wavefront), and another fraction crosses the TR. As the driving perturbation is impulsive, the corresponding wind speed and density response signals have a wide spectral distribution (rather than being monochromatic or having discrete frequencies), and the chromosphere responds to this mechanical perturbation in a frequency-dependent manner. Most of the incident (especially high frequency) wave power effectively crosses the whole chromosphere and leaves the domain through the bottom boundary. The remainder will excite resonant oscillations of the chromosphere, that translate into slight oscillations of the position of the TR and on the excitation of small amplitude secondary compressive waves that propagate upward across the whole corona (as slow-mode waves, such as the stronger initial perturbation). These trailing secondary oscillations are long-lasting (on a time-scale much larger than the 12 minutes of the initial perturbation), but hard to show in the ``x-t'' diagrams above. The first panel of Figure \ref{fig_8} shows them more clearly (e.g. in region B).
It is worth noting that these secondary effects make that the density and temperate profiles do not revert immediately to their initial state (prior to the heating event) after the initial compression and rarefaction fronts (even after the 15 hours of simulation time). Furthermore, the sudden turning on and off of the supplementary heat source, as well as the progressive mass flux increase, change and precondition the background state on which the next perturbation propagates through. As a result, the propagation speeds of similar perturbations change during the course of the simulation, and so do the corresponding slopes in the ``x-t''diagrams. A direct consequence of this is that the full duration of the initial perturbation measured in the corona (even in the low corona, before other physical processes come into play) differs from the initial 12 minutes, and so does the initially expected radial size of the density enhancements in the corona.

\subsection{Periodic heating events on a 1D wind stream}
{\label{sec_results_1Dph}}

\begin{figure}[ht!]
\includegraphics[width=0.48\textwidth]{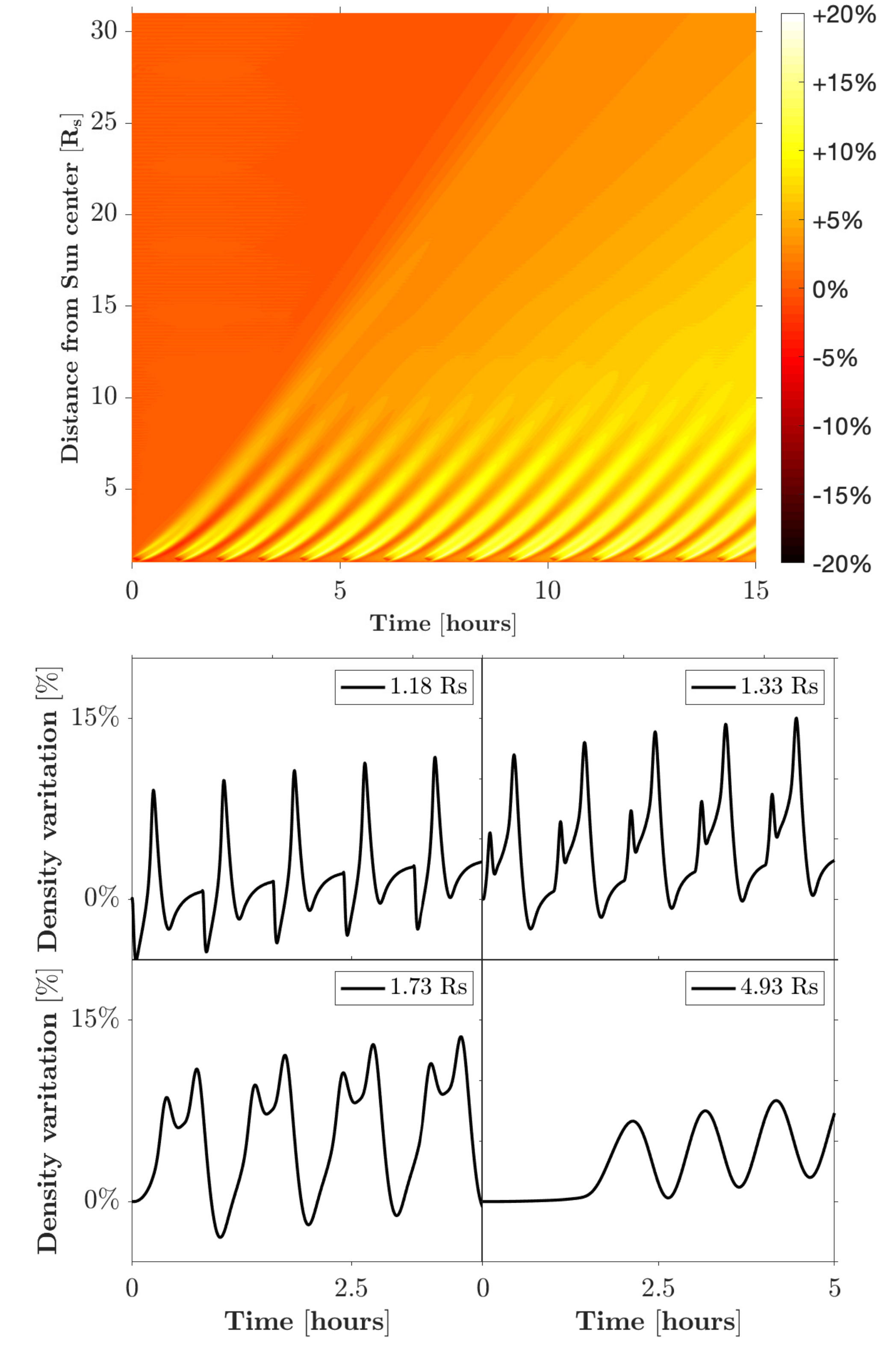}
\caption{{\small \textit{Simulation 2: density variation along the entire grid (from 1 to 31 $R\odot$), over 15 hours. One can notice how the compression fronts pile-up in time, leading to a general density enhancement that is particularly noticeable on the bottom-right graph showing temporal evolution of the density variation at  $4.93 R\odot$.}}} 
\label{fig_5}
\end{figure}

Figure \ref{fig_5} shows a ``x-t'' diagram of the relative base difference of the plasma temperature on a similar case, but this time subject to periodic heating events.
We present and discuss here only one of several simulations ran for different $\delta t_1 $ and $\delta t_2 $ (case 2 of Table \ref{tab:parameters_1D}), as the conclusions can be easily transposed to other cases.

Consecutive compression and rarefaction fronts succeed in direct relation to the specific heating event durations and periodicities, with each one of them evolving in a manner similar to those described in Sect. \ref{sec_results_1Dsh}.
The secondary (and longer-lasting) effects of the heating perturbations can however superpose, unless the time interval between consecutive events is very large. 

The diagram and graphs of the temporal evolution of the density variation at different altitudes clearly show
an increase of the mean density throughout the whole simulation. The mean density increase saturates, though, on a time-scale that depends on the relation of heating periodicity $\delta t_2$ to the duration of the mass evaporation phenomena. In fact, substituting the periodic heating with a steady supplementary heating source with a heat deposition rate equivalent to that of the periodic perturbation produces (asymptotically) similar background / mean density profiles.
In cases with short periodicities, such as the one depicted here (with $\delta t_2 = 1\ h$), consecutive density enhancements coalesce in the high corona. In consequence, the clearly defined compact density structures observed in the high corona should be expected to be produced by coronal heating events with long enough recurrence times, at least above a few hours.

\subsection{Varying the transient heating rate parameters}
{\label{sec_results_ps}}

\begin{figure}[!t]
\includegraphics[scale=0.45]{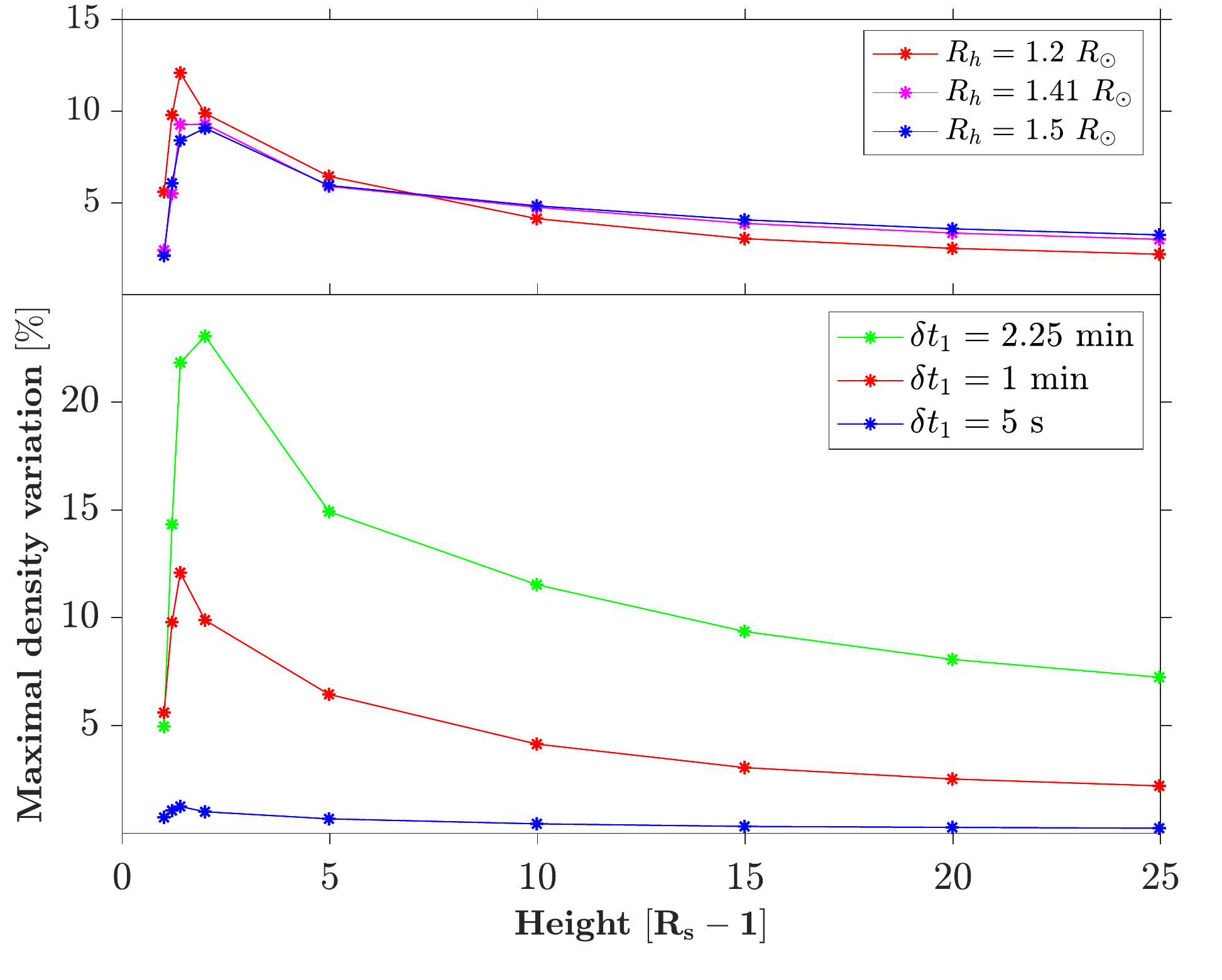}
\caption{{\small \textit{Maximal density variations (over time) measured at different altitudes in 3 simulations (Simulations 3, 4 and 5). The input parameters of those 3 simulations were identical, except for the location of the additional transient duration (top) and for the duration of the transient heating (bottom of Simulations 3, 6 and 7).}}} 
\label{fig_6}
\end{figure}

A parametric study of 1-D simulations of single heating events was run, varying the parameters of the Gaussian function used to model the additional transient heating ($R_h$; see Eq. \ref{eq_heatingrate_exp}).
Figure \ref{fig_6} shows the effect of changing the height of heat deposition and the duration of the heating event in terms of the maximum amplitude of the coronal density enhancements at all coronal altitudes (ensuring that the transient heating region does not overlap with the TR, located at $R_{\rm Tr}=1.008 R\odot$). The relative density variation at the outer corona increases with increasing $R_h$, keeping the amplitude of the transient heating function fixed. This can be understood as the effect of depositing the same amount of energy at places of the corona with different densities; the same energy deposition on a less dense plasma (that is, at a higher altitude) will produce a higher temperature peak (more energy per particle), a stronger relative pressure amplitude, and drive a compression front with a higher density contrast.
Interestingly, this trend reverses in the low corona, with the cases with $R_h > 1.4\ R_{\odot}$ showing a milder density enhancement there. Note that this tentative height threshold corresponds to the temperature peak height in the reference case $R_{\rm Tmax}$, for which we can expect that the fraction of heat that is conducted downward will be lower in the cases with $R_h > R_{\rm Tmax}$. The cases compared here corroborate this idea, as the amplitude of the resulting downward heat flux indeed increases with decreasing $R_h$. Caution should be taken in generalizing this conclusion as the heating event can modify significantly the temperature profile of the low corona (hence potentially displacing $R_{\rm Tmax}$).

Increasing the heating event duration ($\delta t_1$) while keeping all the other heating parameters fixed leads, naturally, to higher density contrasts in the high corona. We only consider here event duration significantly lower than the time-scales of the processes that transport heat across the corona.
The density response to longer $\delta t_1$ necessarily saturates, as that is equivalent to setting an additional but permanent heating source (for which a new stationary state would be reached).

Varying the transient heating function width does not produce meaningfully different density contrasts, as long as the total heat input is kept constant (by normalizing the heating amplitude $a_p$ accordingly).

\subsection{Single heating event on the 3D solar wind}
\label{sec_results_3Dsh}

\begin{figure}[!t]
\includegraphics[width=0.45\textwidth]{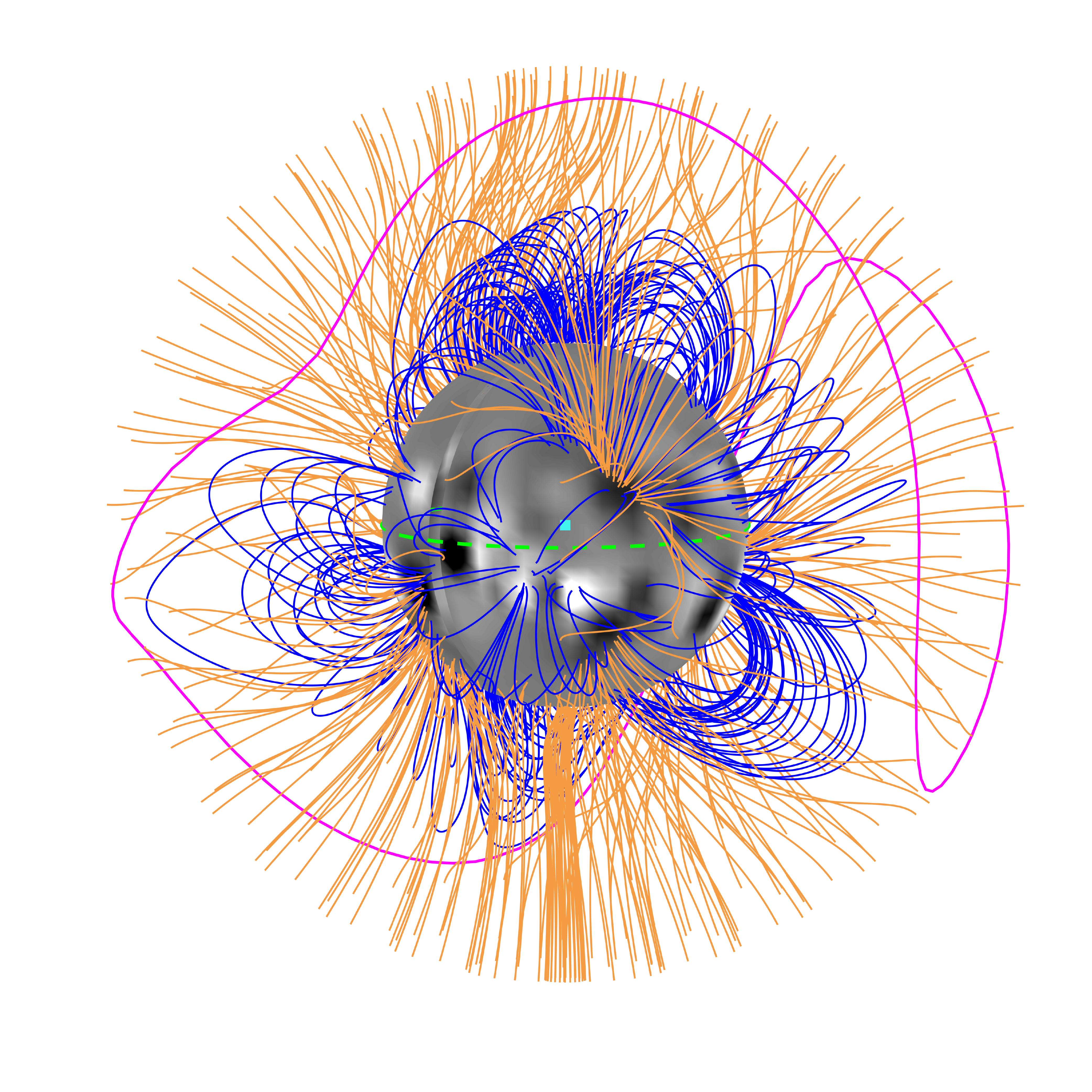}
\caption{{\small \textit{Reconstruction of magnetic field lines during CR 2149 computed using the PFSS method based on a WSO synoptic map of the surface magnetic field. Orange and blue lines represent open and closed fieldlines. The HCS is plotted in violet, the ecliptic is represented by a dashed green line, and the cyan marker indicates the position of STEREO-A on April 15, 2014.}}} 
\label{fig_7}
\end{figure}

The STEREO deep-field campaign described in \citet{Deforest2018} was held at around April 14, 2014, during a solar activity maximum. Figure \ref{fig_7} shows the three-dimensional topology of the magnetic field in the low corona, computed as a Potential Field Source Surface (PFSS) extrapolation based on a synoptic magnetogram from the Wilcox Solar Observatory (WSO) for Carrington rotation 2149. Closed magnetic loops are traced as blue lines, while magnetic field lines that open up to the interplanetary space are traced in orange.
The heliospheric current sheet (HCS) displays a particularly complex configuration (violet line), appearing side-on almost everywhere from the vantage point of STEREO-A (the position of the spacecraft is marked in cyan over the Sun). The dashed green line represents the solar equatorial plane.
Coronal streamers and pseudo-streamers appear at all latitudes, the coronal holes cover a small surface area, and the open magnetic flux tubes show a wide range of areal expansions and deviations from the vertical direction \citep[letting us guess a complex solar wind structure in result; \emph{cf.}][]{pinto_flux-tube_2016}.
We traced out a large sample of open magnetic field lines covering the whole open corona, and computed a full solar wind profile for each one of them using MULTI-VP. The plasma heating parameters are set up exactly as in Simulation 1, the only difference 
being that there are now many contiguous flux-tubes magnetic flux tubes whose geometries (magnetic field amplitude, orientation, expansion) are obtained from the PFSS extrapolation shown in Figure \ref{fig_7}. 
The wind profiles obtained are then remapped into a three-dimensional spherical volume. 
We obtained first a global steady-state solar wind solution (in the same general manner as in the 1-D cases in the previous sections) that we used as initial condition for the simulations that followed.
The first three-dimensional case we discuss here (simulation 8) is the direct counterpart of our 1-D case 1 (see Sect. \ref{sec_results_1Dsh} and Table \ref{tab:parameters_1D}). We applied a single transient heating event with exactly the same parameters for $R_{\rm p}$, $r_{\rm p}$ and $a_{\rm p}$, but this time on a complex and realistic magnetic field configuration that drives a very structured background solar wind.

\begin{figure*}[!t]
  \centering
  \includegraphics[width=.9\textwidth]{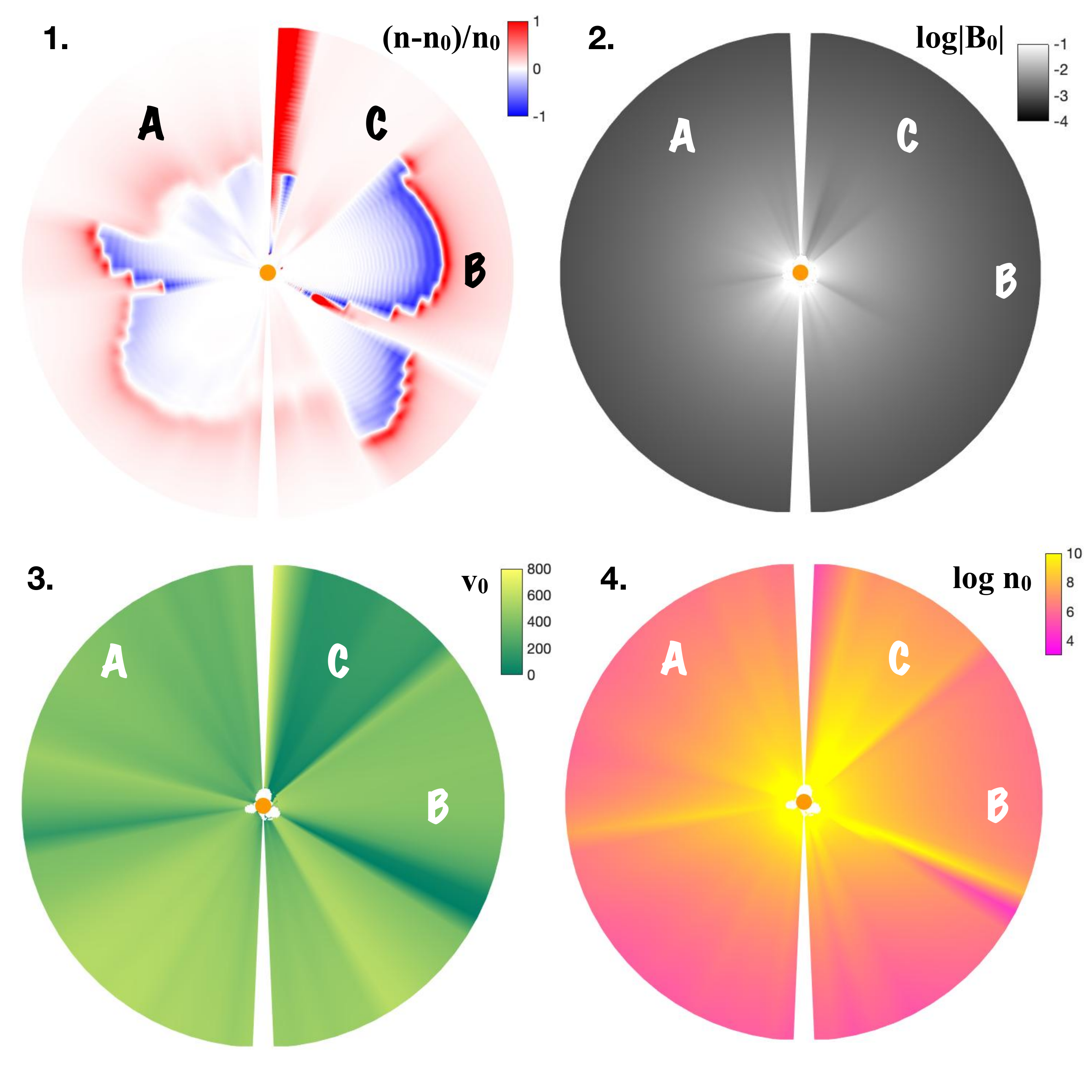}
  \caption{{\small \textit{Simulation 8: Four slices of the 3D simulation domain corresponding to simulation 8, in the plane of sight of STEREO-A on April 15, 2014. On the top-left slice (labelled 1.), the density variation $(n-n_0)/n_0$ is plotter in red-to-blue colour from 1 to -1, at 7.5 hours after the single transient heating is added to the steady state solution. The steady state solution profiles of magnetic field (top-right, number 2.), solar wind velocity (bottom-left, number 3.) and density (bottom-right, number 4.) are also displayed. Magnetic field is given in $nT$, velocity in $km/s$ and density in $cm^{-3}$}. Magnetic field and density are given in log scale. On each slice, three regions of particular interest are labelled A, B, and C.}} 
  \label{fig_8}
\end{figure*}

Figure \ref{fig_8} shows four slices of the 3D simulation domain, corresponding to the plane-of-sky as seen by STEREO-A on April 15, 2014.
The first panel shows the relative variation of the wind density in respect to the initial background state (relative base difference) at an intermediate phase of the propagation of the initial compression front. The other panels show the magnetic field amplitude, wind speed and density in the same planes for the background wind state (the initial state).
Some regions of particular interest are denoted A, B and C.
Regions A and B correspond to mildly fast solar wind zones (with B being a more typical fast wind flow, with a higher terminal wind speed and lower density). Region C is composed of very slow and dense solar wind streams that are formed in thin coronal holes that surround a large pseudo-streamer (placed close to the north / north-western portion of the visible solar limb in Fig. \ref{fig_7}), and that evolve through a much more complex magnetic configuration in the low corona.
It is readily visible that the propagation of the initial compression front is extremely anisotropic, reflecting the large variety of specific magnetic flux tube geometries and wind streams properties found at each position around the Sun (in respect to a heating perturbation introduced everywhere at the same height).
The very non-spherical shape of the compression front traces directly the distribution of the slow-mode propagation speed (slow mode phase speed plus wind speed component in the plane-of-sky). The perturbation propagates faster in region B than in region A, and much slower in region C.
The relative amplitude of the main compression front also varies significantly, being inversely related to the density of the background flow at the position at which the transient heating source was added (in agreement with the conclusions of Sect. \ref{sec_results_ps} for the 1-D cases).
The density fluctuations that trail the first compression front correspond to (numerically resolved) small amplitude slow-modes that are triggered by resonant oscillations of the chromosphere in response to the initial pressure pulse, as discussed in Sect. \ref{sec_results_1Dsh}. These seem to be more preponderant in clear fast wind flows such as those in region B (but they do occur in most other places).
The heating event provokes a slower mass-flux build-up into the corona (evaporation), that is more preponderant on wind streams with temperature profiles that favour a stronger downward conductive heat flux.
Some wind streams sometimes show very mild compression fronts, and the base difference images (first panel in Fig. \ref{fig_8}) will show a ray-like feature that corresponds to a progressive mass loading of the corresponding flux tube.
The relative preponderance of all of these different contributions on the solar wind density fluctuations hence depend strongly on the specific conditions of each stream (magnetic geometry, plasma stratification, thermo-dynamical properties of the wind flow), and even very simple uniform perturbation to the background heating in the low corona can lead to rather diverse and complex signatures in the high corona.

The meridional slices in Fig. \ref{fig_8} highlight sharp gradients between different regions of the solar wind (\emph{e.g.} the boundary between regions B and C), as these regions are complex three-dimensional structures that cross the slice. These boundaries would appear smoothed out in white-light imagery of the real solar wind, as emission the optically thin emission from the background and from foreground of the plane-of-sky would superpose.

\subsection{Recurrent heating events on the 3D solar wind}
{\label{sec_results_3Drph}}

Heating perturbations occurring in the low corona (related tentatively to CBPs) should be expected to be triggered in a much more random manner than in the simulations we described in the previous section.
For that reason, we ran another three-dimensional simulation on which, unlike simulation 8 in the previous section, we have randomized the parameters of the heating functions (simulation 10 in Table \ref{tab:parameters_1D}).
The transient heating function on Eq. \eqref{eq_heatingrate_exp} was assigned one set of parameters for each individual wind stream simulated, such that $\delta t_1$ and $\delta t_2$ are set differently (and randomly) for each flux tube. A random phase (\emph{i.e.} the time left for the next heating event) was also introduced. Taking into account that on mid-April 2014 there were approximately 550 Coronal Bright Points measured daily on the visible solar disk \citep{Alipour2015},
and that we used an ensemble of open magnetic flux tubes that cover the outer corona with an angular resolution of $5\times 5^\circ$ (coherent with resolution of the WSO magnetogram used),
we defined the time interval between consecutive heating events to be $\delta t_2 \approx 30\ \mathsf{h}$ (distributed uniformly in an interval between $26.25$ and $33.75 \mathsf{h}$).
Following \cite{Alipour2015}, we set the mean duration of a transient $\delta t_1$ to be of about 12 minutes (uniformly distributed between approximately $11$ and $13$ minutes. We did not attempt at implementing more sophisticated statistics (e.g. with heating duration anti-correlated with heating frequency) on this manuscript.
All the other parameters defining the transient heating function are similar to those in Simulation 1. We ran simulation 10 for about 30 hours of physical time.

\begin{figure}[!t]
  \centering
  \includegraphics[width=\linewidth]{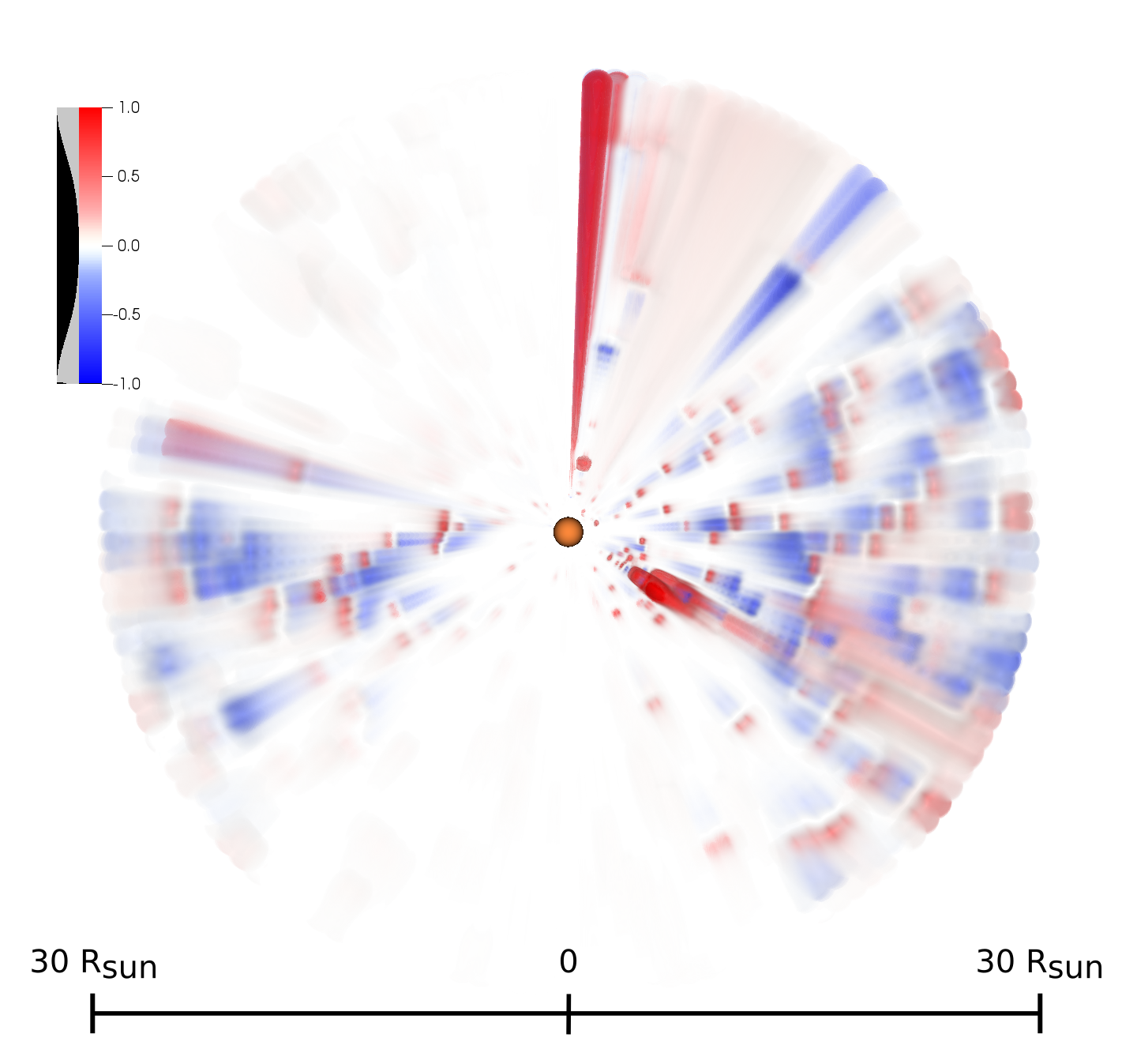}
  \caption{{\small \textit{Simulation 9: Volume rendering of the relative density fluctuations in the case with random heating events (at time $t=25.6\mathrm{h}$ in respect to $t=0$ in our simulation 10) from the point of view of STEREO-A, including $\pm 20^\circ$ in front and behind of the plane of sky shown in Fig. \ref{fig_8}. The orange sphere represents the Sun, and the axis indicates the size of the domain (that extends up to 30 $R_{\odot}$. Red and blue colours indicate respectively density increases and decreases. Null differences are rendered transparent.
        The superposition of density perturbations propagating in the foreground and background is evident.}}}
  \label{fig_9}
\end{figure}

Figure \ref{fig_9} shows the relative density variations found in the simulation with respect to the unperturbed state within a volume defined within the longitudes $\pm 20^\circ$ around the image plane that would be visible from the position of STEREO-A on April 2014.
The image was built using a volume rendering technique, on which each volume element of the numerical grid (or voxel) was assigned a colour and an opacity as a function of its value.
We set red and blue tones for positive and negative density variations, respectively, and we assigned an opacity table that made the volume elements having $\sim 0$ density variations be transparent, and those having strong density perturbations be opaque. 
The interest of this representation lies in showing the back/foreground superposition as perceived from a given point of view, unlike in Fig. \ref{fig_8}.
The instant represented corresponds to $t=25.6\ h$ after the beginning of the simulation.
During the initial phases (first hours) of the simulation, the propagation of elemental compressive fronts follows the general progression displayed in Fig. \ref{fig_8} (e.g, with perturbations in region B travelling faster across the plane-of-sky than perturbations on regions A and C).
Once the simulation has run for long enough time for the perturbation patterns to settle in on all flux tubes, 
compact density fluctuations become ubiquitous (although with a higher prevalence in regions with faster and less dense winds).
The superposition of foreground and background density structures results in a very rich fluctuation pattern, despite the seemingly weak statistics for the recurrence of CBPs on individual wind streams.

\begin{figure*}[!t]
  \centering
  \includegraphics[width=\linewidth]{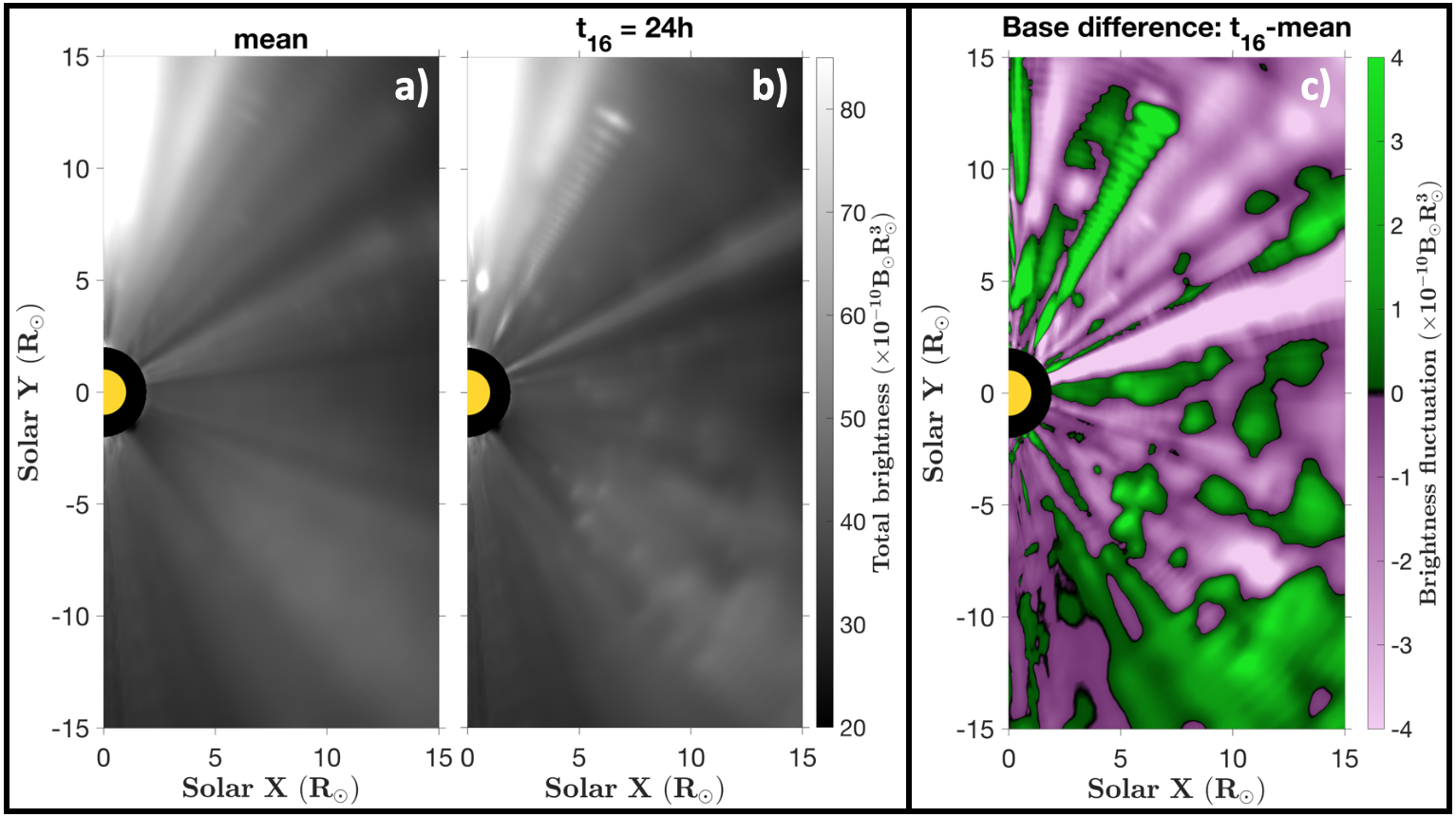}
  \caption{{\small \textit{
        Panel a: synthetic white-light image averaged over times $t=15\mathrm{h}$ to $t=30\mathrm{h}$ from Simulation 9. 
        Panel b: synthetic image at time $t=24\mathrm{h}$.
        Panel c: base difference image at time $t=24\mathrm{h}$ with respect to the mean background image (panel a).}}} 
  \label{fig_10}
\end{figure*}

Figure~\ref{fig_10} shows a series of synthetic white-light images computed from the simulation results.
The goal is to estimate the white-light fluctuations induced by the simulated density perturbations as they would be detected by the STEREO-A COR2 coronograph.
We proceeded by first defining a number of points along each line-of-sight of the instrument, which are fetched from the SolarSoft package in IDL. Then, for all those points, we compute the total brightness following the Thomson scattering theory \citep[][]{Howard2009Ssr}. Given a specific location in space and an electron density, this theory provides the amount of white-light photospheric light that is scattered toward the observer by the electrons. Finally, we sum up all the brightness contributions along each line-of-sight of the instrument to produce a synthetic white-light image. This procedure is further detailed in \citet{Poirier2020ApJS}, where synthetic images were produced to interpret the morphology of streamer rays as observed in white-light by the heliospheric imager (WISPR), aboard \textit{Parker Solar Probe}.
Figure~\ref{fig_10}b shows a STEREO COR2 synthetic plain image at time $t=24\mathrm{h}$ computed from the density output of Simulation 10 (only half of the image is shown). As done in \citet{Deforest2018} the brightness is multiplied by $r^3$ (where $r$ is the heliocentric distance) to enhance the visibility of the fluctuations. We calculated a background image that is an average of all synthetic images within a specific time interval (see panel a). This interval starts at $t=15\mathrm{h}$, after the first perturbations cross the whole domain. Figure~\ref{fig_10}c is a base difference image (between the images in panels a and b) that highlights brightness fluctuations at a specific time with respect to an average state of Simulation 10. Figure~\ref{fig_10}c shows several propagating structures characterised by increases in total brightness $\lesssim4\times10^{-10}\ B_{\odot}R_{\odot}^3$ or alternatively $\sim11\%$ with respect to the mean background image. The shape and order of magnitude are consistent with the ubiquitous brightness fluctuations observed by STEREO COR2 and analysed in \citet{Deforest2018}. Synthetic brightness fluctuations show different shapes, scales and intensities. Regarding shapes, one can see patches (or blobs) alongside elongated features. Elongated features are mainly visible due to the saturation of the colour scheme, where neighbouring fluctuations coalesce. They may also result from the superposition of background and foreground contributions.

\section{Discussion and conclusions}
We have investigated the effects of transient coronal heating events on the solar wind by means of numerical simulations. 
We attributed these heating events to CBPs that appear inside coronal holes, and verified that the observed occurrence rate, mean duration, height of occurrences and temperature \citep{Alipour2015}, 
were used as a reference to define the properties of heat deposition events at the coronal base of open magnetic flux tubes. 
The transient heating events generate compression fronts that propagate outward as slow-mode waves up to several tens of solar radii. The perturbations (and the corresponding density enhancements) travel at the local sound speed in the moving frame of the solar wind plasma. 
The heating event also generates a compression front that propagates downward into the chromosphere, where it excites low amplitude resonant oscillations.
The transition region is also perturbed by the downward conductive heat flux that results in a chromospheric evaporation episode on a time scale much longer than the duration of the transient heating event. The added mass loading gradually alters the mean background state of the wind, making it denser and less fast. After one single heating event of 12 minutes, it can take more than 30 hours for the corona to retrieve its initial state. Therefore periodic heating events separated by less than a day can have a durable effect on the average coronal density. 
Heat conduction also quickly transports heat upward from the perturbed region.
Each one of these different effects is more or less preponderant according to the altitude of heat deposition in respect to the positions of the temperature peak and of the sonic point on each individual wind stream.
The duration of the heating event also impacts the amplitude of the density variation induced in the corona. \\

We ran time-dependent solar wind simulations on thousands of magnetic field lines spanning the whole three-dimensional corona in order to analyse the morphology of the out-flowing perturbations. 
Statistically, the long duration between consecutive heating events on a single field line (of more than a day) means that a unique density pulse propagates inside the COR-2 field of view along each simulated flux tube. 
Since magnetic flux tubes are radial at these heights, this means a single flux tube can contribute with only one density pulse along a particular position-angle (PA). The modelling of thousands of tubes combined with the effect of line of sight integration shows that numerous density peaks can be visible simultaneously in the solar corona along a fixed PA. Synthetic white-light images produced from the simulated density cubes reveal that the induced brightness variations stand out particularly well inside high-speed streams that are generally of lower density and therefore less bright. Considering the entire corona, the simulations show that at the heights imaged by COR-2 the brightness variations are ubiquitous and comparable with the images shown in \citet{Deforest2018}. 
Our study reveals that at least a fraction of the brightness fluctuations observed in the corona can be caused by heating events due to CBP occurring below, near the coronal base.
Our goal was not to reproduce the exact variability observed by \citet{Deforest2018}, but rather to check whether CBPs could be the source of (at least part of) the observed density structures.
We restricted our study to a simplified set of temporal lengths and spatial distribution of CBPs inspired by the statistical distributions provided by \citet{Alipour2015}. 
Detailed statistics describing CBP frequency as a function of heating amplitude, as well as detailed temporal profiles for the heating events were left for future consideration.
Moreover, our simulations only consider open magnetic flux tubes, and all CBPs are not observed above coronal holes. For this reason, the number of density variations present in the simulation results of section \ref{sec_results_3Drph} can be over-estimated.
The transient heating produced via magnetic reconnection is maybe not as intense as what is set in the simulation, and may also be shorter than the duration of the CBP itself. 

The interpretation that density enhancements in the solar wind originate in CBPs rests on the interpretation that CBPs involve magnetic reconnection with the open magnetic field. Recently, \citet{Raouafi2014} studied the presence of Coronal Bright Points — or, as they call them, ‘Plume Transient Bright Points (PTBPs)’ — above equatorial coronal holes and explained how the presence of those extremely hot, frequent and short ($\sim$10 minutes) areas at the footpoints of open magnetic flux tubes could play a key role in the formation and sustainability of coronal plumes. \cite{Wang1998} also showed that jets are observed in white-light and EUV images above coronal holes. Our main conclusion from the periodic transient heating in section \ref{sec_results_1Dph} support this assumption: periodic short and intense heating can lead to a sustained enhancement of plasma density that could contribute to the development of coronal plumes. \cite{Raouafi2014} (and references therein) also underlined that the intensity and duration of PTBPs could be directly related to the increase or decrease of local magnetic reconnection rates, and that CBPs are ubiquitous and randomly spatially spread on the visible solar disk whatever solar activity level. The present study is compatible with such a statement, showing that transient heating added randomly in space but dependent on local magnetic field intensity can lead to ubiquitous density variations, and thus ubiquitous brightness fluctuations in white-light images. 

We did not take into account the magnetic field re-configurations that are likely to occur during bright points: our model only responds to the associated impulsive heating. Furthermore, we do not consider cross-field propagation, meaning that all energy input is channelled in the direction parallel to the field (via sound/slow-mode waves, thermal conduction, advection, and neglecting e.g, kink/fast modes).
CBPs can result from magnetic reconnection induced by the magnetic element that emerges inside the unipolar open field of coronal holes \citep{Wang1998}. Magnetic reconnection should create kinked field lines propagating outward along the open field and perhaps also contributing to the compression of the plasma in the jet \citep{Roberts2018}.
Parker Solar Probe has recently revealed that large folds in the magnetic fields are ubiquitous in the solar wind \citep{Bale2019, Kasper2019}. While the folds measured in the solar wind from coronal holes are not associated with significant density fluctuations, those originating inside streamer flows have been shown to be associated with strong density variations \citep{Rouillard2020ApJS}. It is yet unclear where these folds and density variations originate. We will address this mystery in a future study that combines full 3-D MHD simulations with remote-sensing and solar wind measurements near the Sun.

\acknowledgments
The numerical simulations of this study were performed using HPC resources from CALMIP (Grant P1504). The IRAP team acknowledges support from the French space agency (Centre National des Etudes Spatiales; CNES; \url{https://cnes.fr/fr}) that funds activity in plasma physics data center (Centre de Données de la Physique des Plasmas; CDPP; \url{http://cdpp.eu/}) and the Multi Experiment Data \& Operation Center (MEDOC; \url{https://idoc.ias.u-psud.fr/MEDOC}),  and the space weather team in Toulouse (Solar-Terrestrial Observations and Modelling Service; STORMS; \url{https://stormsweb.irap.omp.eu/}). This includes funding for the data mining tools AMDA (\url{http://amda.cdpp.eu/}), CLWEB (\url{clweb.cesr.fr/}) and the propagation tool (\url{http://propagationtool.cdpp.eu}).  The work of L. Griton, A.P. Rouillard and N. Poirier was funded by the ERC SLOW{\_}\,SOURCE project (SLOW{\_}\,SOURCE - DLV-819189). A. Kouloumvakos and R. Pinto also acknowledge financial support from the ANR COROSHOCK (ANR-17-CE31-0006-01) and FP7 HELCATS projects \url{https://www.helcats-fp7.eu/}.

\bibliographystyle{apj}
\bibliography{transient_heating_biblio}
%
\end{document}